\documentclass[12pt]{JHEP3}

\usepackage{amsmath}

\newcommand{\e}{\epsilon}

\renewcommand{\L}{{\mathcal{L}}}
\newcommand{\bL}{\bar{{\mathcal{L}}}}
\newcommand{\z}{{\bar z}}

\renewcommand\O{{\mathcal{O}}}
\newcommand{\be}[1]{ \begin{equation}\label{#1} }
\newcommand{\ee}{\end{equation}}
\newcommand{\ben}[1]{\begin{eqnarray}\label{#1} }
\newcommand{\een}{\end{eqnarray}}
\newcommand{\eq}[1]{(\ref{#1})}
\newcommand{\p}{\partial}

\newcommand{\cP}{\mathcal{P}}
\newcommand{\cJ}{\mathcal{J}}
\newcommand{\refb}[1]{(\ref{#1})}

\title{The BMS/GCA correspondence}

\author{
Arjun Bagchi
$^{1,2}$\\
$^1$Harish-Chandra Research Institute,\\
$\,$Chhatnag Road, Jhusi,\\
$\,$Allahabad 211019, India\\

$\;$ $^2$Institute for Advanced Study,\\
$\;$ $\,$Princeton, NJ 08540, USA\\

$\;$\email{arjun@hri.res.in}
}

\abstract{We find a surprising connection between asymptotically flat space-times and non-relativistic conformal systems in one lower dimension. The BMS group is the group of asymptotic isometries of flat Minkowski space at null infinity. This is known to be infinite dimensional in three and four dimensions. We show that the BMS algebra in 3 dimensions is the same as the 2D Galilean Conformal Algebra which is of relevance to non-relativistic conformal symmetries. We further justify our proposal by looking at a Penrose limit of a radially infalling null ray inspired by non-relativistic scaling and obtain a flat metric. The 4D BMS algebra is also discussed and found to be the same as another class of GCA, called the semi-GCA, in three dimensions. We propose a general BMS/GCA correspondence. Some consequences are discussed.}

\begin{document}

\baselineskip 3.5ex

\section{Introduction}

Holography in asymptotically Anti de Sitter spaces has been the cynosure of attention for over a decade, following the AdS/CFT correspondence \cite{Maldacena:1997re}. Somewhat less studied and even lesser understood is holography in asymptotically flat spacetimes \cite{Susskind:1998vk}. One of the approaches to this has been to consider the Bondi-Metzner-Sachs group. In the absence of gravity, the isometry group of the space-time is the well know Poincare group which is the semi-direct product of translations and Lorentz transformations. The situation, however, changes drastically when gravity is turned on, even for weak gravitational fields. When one looks at four dimensional asymptotically flat metrics, the isometry group of the background metric is enhanced to an infinite dimensional asymptotic symmetry group at null infinity. This is the Bondi-Metzner-Sachs group \cite{BMS}. This consists of the semi-direct product of the global conformal group of the unit 2-sphere and the infinite dimensional ``super-translations''. There is a further enhancement to two copies of the Witt algebra (Virasoro algebra without the central extension) times the super-translations if one does not require the transformations generated to be well defined \cite{Barnich:2010eb}. 
In three dimensions, under similar considerations, $\mathfrak{bms_3}$ is again infinite dimensional and now has one copy of the Witt algebra along with the supertranslations \cite{Barnich:2006av}. These correspond to a null boundary of $S^1 \times R$. For a more thorough review of the BMS group with a modern perspective, the reader is referred to \cite{Arcioni1}. Other studies of the flat space holographic correspondence using the BMS group can be found in \cite{ref1}. An incomplete list of works on other aspects of flat space holography include \cite{refer2}. 

The Galilean Conformal Algebra (GCA) on the other hand, has been discussed in literature in connection with a non-relativistic limit of the AdS/CFT conjecture \cite{Bagchi:2009my}. It was obtained by a parametric contraction of the finite conformal algebra and was observed to have an infinite dimensional lift for all space-time dimensions. The 2 and 3 point correlation functions of the GCA were found \cite{Bagchi:2009ca}. In the case of two dimensions, the relativistic algebra is itself infinite dimensional and a simple map was obtained between the two copies Virasoro algebra and 2d GCA. The representations and other details of the two dimensional case have been worked out in detail in \cite{Bagchi:2009}. For other related studies in the GCA, the reader is referred to \cite{Duval:2009vt} -- \cite{Hotta:2010qi}.

In what follows, we briefly review the two algebras the BMS{\footnote{The conventions and indeed the review of the BMS algebra closely follow \cite{Barnich:2006av}.}} and the GCA in arbitrary dimensions and then show that $\mathfrak{bms}_3$ and $\mathfrak{gca}_2$ algebras are isomorphic. The realization of central charges are discussed. We comment on how the limiting procedure from the asymptotically AdS to the BMS as discussed in \cite{Barnich:2006av}, although similar, is a bit different from the way the GCA is obtained from the relativistic Virasoro in \cite{Bagchi:2009}. In \cite{Bagchi:2009my}, we proposed a non-relativistic scaling for the bulk theory. Here we show that a similar scaling in $AdS_3$ would lead us to a flat metric when one performs a Penrose limit on a radially infalling null ray. 
We move on to $\mathfrak{bms_4}$ and discuss the equivalence to another algebra, the semi-Galilean conformal algebra \cite{Alishahiha:2009np}, in three dimensions. The semi-GCA is obtained by a generalized contraction which makes one of the spatial directions non-relativistic while keeping the other relativistic. A general BMS/GCA correspondence is proposed along with various comments.

\pagebreak

\section{A quick look at the two algebras}

\subsection{The $\mathfrak{gca_n}$ algebra}

The maximal set of conformal isometries of Galilean spacetime generates the infinite dimensional Galilean Conformal
Algebra \cite{Bagchi:2009my}. The notion of Galilean spacetime is a little subtle since the spacetime metric degenerates 
into a spatial part and a temporal piece. Nevertheless there is a definite limiting sense (of the relativistic spacetime) in which one can define the conformal isometries (see \cite{Duval:2009vt}) of the nonrelativistic geometry. Algebraically, the set of vector fields generating these symmetries 
are given by
\ben{gcavec}
L^{(n)} &=& -(n+1)t^nx_i\p_i -t^{n+1}\p_t \,,\cr
M_i^{(n)} &=& t^{n+1}\p_i\,, \cr
J_a^{(n)} \equiv J_{ij}^{(n)} &= & -t^n(x_i\p_j-x_j\p_i)\,,
\een 
for integer values of $n$. Here $i=1\ldots (d-1)$ range over the spatial directions. 
These vector fields obey the algebra
\ben{vkmalg}
[L^{(m)}, L^{(n)}] &=& (m-n)L^{(m+n)}, \qquad [L^{(m)}, J_{a}^{(n)}] = -n J_{a}^{(m+n)}, \cr
[J_a^{(n)}, J_b^{(m)}]&=& f_{abc}J_c^{(n+m)}, \qquad  [L^{(m)}, M_i^{(n)}] =(m-n)M_i^{(m+n)}. 
\een

There is a  finite dimensional subalgebra  of the GCA (also sometimes
referred to as the GCA) which consists of taking $n=0,\pm1$ for the
$L^{(n)}, M_i^{(n)}$ together with $J_a^{(0)}$. This algebra is obtained
by considering the nonrelativistic  contraction of the usual (finite
dimensional) global conformal algebra $SO(d,2)$ (in $d>2$ spacetime
dimensions) 
(see for example \cite{negro1997}--\cite{Bagchi:2009my}). 

However, in two spacetime dimensions, as is well known, the situation is special. The relativistic conformal algebra is infinite dimensional and consists of two copies of the Virasoro algebra. One expects this to be also related, now to the infinite dimensional GCA algebra. Indeed in two dimensions the non-trivial generators in  \eq{vkmalg} are the $L_n$ and the $M_n$ (where we have dropped the spatial index from the latter since there is only one spatial direction and instead restored the mode number $n$ to the conventional subscript) :
\ben{gca2dvec}
L_n &=& -(n+1)t^n x\p_x -t^{n+1}\p_t\,, \cr
M_n &=& t^{n+1}\p_x\,,
\een
which obey
\ben{vkmalg2d}
[L_m, L_n] &=& (m-n)L_{m+n}\,, \qquad [M_{m}, M_{n}] =0\,, \cr
[L_{m}, M_{n}] &=& (m-n)M_{m+n}\,. 
\een

The generators  in \eq{gca2dvec} arise precisely from a nonrelativistic contraction of the two copies of the Virasoro algebra of the relativistic theory. The non-relativistic contraction consists of taking the scaling 
\be{nrelscal}
t \rightarrow t\,, \qquad   x \rightarrow \epsilon x\,,
\ee
with $\epsilon \rightarrow 0$. This is equivalent to taking the velocities $v \sim \epsilon$ to zero
(in units where $c=1$).

Consider the vector fields which generate (two copies of) the centre-less Virasoro Algebra in two dimensions :
\be{repn2dV}
\L_n = -z^{n+1} \p_z\,, \quad \bL_n = -\z^{n+1} \p_{\z}\,.
\ee 
In terms of space and time coordinates, $z= t+x$, $\z=t-x$. Hence $\p_z = {1\over 2} (\p_t + \p_x)$ and $\p_{\z} = {1\over 2} (\p_t - \p_x)$. Expressing $\L_n , \bL_n$ in terms of $t,x$ and taking the above scaling  \eq{nrelscal} reveals that in the limit the combinations
 \ben{GCArepn}
\L_n + \bL_n &=& -t^{n+1}\p_t - (n+1)t^nx \p_x + \O(\e^2) \,,\cr
\L_n - \bL_n &=& -{1\over \e}t^{n+1} \p_x + \O(\e)\,.
\een 
Therefore we see that as $\e\rightarrow 0$
\be{Vir2GCA}
\L_n + \bL_n \longrightarrow L_{n}\,, \quad \e (\L_n - \bL_n) \longrightarrow - M_{n}\,.
\ee

Thus the GCA in 2d arises as the non-relativistic limit of the relativistic algebra.

\subsection{The $\mathfrak{bms_n}$ algebra}

In this subsection, we look first at the algebra of the BMS group in general dimensions and then focus on the three-dimensional case. We wish to recover the BMS group from the symmetries of flat space-time. To find the asymptotic symmetries we would need to look at the structure at null infinity. Let us begin by introducing the retarded time $u=t-r$, the luminosity distance $r$ and
angles $\theta^A$ on the $n-2$ sphere by $x^1= r\cos \theta^1$, $x^A =
r\sin \theta^1\dots\sin \theta^{A-1}\cos \theta^A$, for
$A=2,\dots,n-2$, and $x^{n-1}=r\sin \theta^1\dots \sin \theta^{n-2}$. The Minkowski metric is then given by
\be{metric_flat}
d\bar s^2=-du^2-2dudr +r^2\sum_{A=1}^{n-2}s_A(d\theta^A)^2,
\ee
where $s_1 = 1$, $s_A = \sin^2 \theta^1\dots \sin^2\theta^{A-1}$ for
$2\leq A\leq n-2$. The (future) null boundary is defined by $r =\mbox{constant} \rightarrow \infty$ with $u,\theta^A$
held fixed.

One requires asymptotic Killing vectors to satisfy the Killing equation to leading order. They turn out to be \cite{Barnich:2006av}
\ben{asykvf}
\xi^u = T(\theta^A)+u\p_{1}Y^1(\theta^A)+o(r^0),\qquad \xi^r= -r\p_{1} Y^1(\theta^A)+o(r),\\
\xi^A = Y^A(\theta^B)+o(r^0),\qquad A=1\dots n-2.
\een
where $T(\theta^A)$ is an arbitrary function on the $n-2$ sphere, and $Y^A(\theta^A)$ are the components of the conformal Killing
vectors on the $n-2$ sphere. These asymptotic Killing vectors form a
sub-algebra of the Lie algebra of vector fields and the bracket
induced by the Lie bracket $\hat \xi=[\xi,\xi^\prime]$ is determined by
\ben{bracket}
\hat T= Y^A\p_{A} T^\prime + T \p_{1} Y^{\prime 1} - Y^{\prime
A}\p_{A} T -T^\prime \p_{1} Y^{ 1}\,, \\
\hat Y^A = Y^B\p_{B} Y^{\prime A} -Y^{\prime B}\p_{B} Y^{A}.
\een
The asymptotic Killing vectors with $T=0=Y^A$ form an abelian subalgebra in the algebra of asymptotic Killing vectors. The quotient
algebra is defined to be $\mathfrak{bms_n}$.  It is the semi-direct sum of the conformal Killing vectors $Y^A$ of Euclidean $n-2$
dimensional space with an abelian ideal of so-called infinitesimal supertranslations. 

In three dimensions, the conformal Killing equation on the circle 
imposes no restrictions on the function $Y(\theta)$. Therefore,
$\mathfrak{bms_3}$ is characterized by $2$ arbitrary functions
$T(\theta),Y(\theta)$ on the circle. These functions can be Fourier
analyzed by defining $P_n \equiv \xi(T=exp{(i n \theta)},\,\ Y=0)$ and
$J_n = \xi(T=0,\,\ Y=\exp{(in\theta)})$. In terms of these generators, the
commutation relations of $\mathfrak{bms_3}$ become{\footnote{Here we drop all factors of $i$.}}
\be{alg_bms}
[ J_m,J_n] = (m-n) J_{m+n},\qquad  [ P_m,P_n] =0, \qquad  [J_m,P_n] = (m-n) P_{m+n}.
\ee
The 6 dimensional Poincar\'e algebra $\mathfrak{iso(2,1)}$ of 3 dimensional Minkowski spacetime is enhanced
to the semi-direct sum of the infinitesimal diffeomorphisms on the circle with the infinitesimal supertranslations.

\vfill

\section{The $\mathfrak{bms_3 / gca_2}$ correspondence}

From \refb{alg_bms} and \refb{vkmalg2d}, it is obvious that the algebras are isomorphic with the trivial identifications
\be{idf}
L_n \leftrightarrow J_n, \quad M_n \leftrightarrow P_n
\ee 
So, what we have is a holographic correspondence between an asymptotic 3 dimensional flat space-time and a 2 dimensional non-relativistic conformal field theory. 

There have been some different forms of realizations of the bulk theory which has the GCA as its boundary algebra. 
Originally in \cite{Bagchi:2009my}, we proposed the dual gravity theory to be a Newton-Cartan like $AdS_2 \times R^d$ by taking a similar non-relativistic limit on the bulk Anti de Sitter space. So, for the case of $AdS_3$, which is the focal point of our attention now, the theory in the bulk was a $AdS_2 \times R$ Newton-Cartan. The $L_n$'s turned out to be the asymptotic symmetries in the sense of Brown and Henneaux \cite{Brown:1986nw}. 
It has also been observed that the GCA emerges as the asymptotic symmetry algebra of Cosmological Topologically Massive Gravity in three dimensions when the coefficient of the gravitational Chern-Simons term is made very large \cite{Hotta:2010qi}. This realization of the GCA in the bulk allows for an asymmetry which is required in the central charges $C_1$ and $C_2$ discussed below. 
The new gravity description in terms of the BMS algebra gives a third and possibly the most intriguing occurrence of the GCA. 

Some might argue that in order to really have this correspondence, we would need a concrete realization of the boundary theory. To answer this question it is to be noted that the infinite 2d GCA also makes its appearance in non-equilibrium statistical mechanical systems \cite{Henkel06}. So we indeed do have realizations of this BMS/GCA correspondence. 

The Euler equation in non-relativistic hydrodynamics emerges in situations when the viscosity of the fluid is negligible. In \cite{Bagchi:2009my}, it was noted that the finite dimensional GCA is the symmetry algebra of the Euler equations. In fact, it is interesting that all the $M_n$'s (for any $n$) are also symmetries of the equations \cite{Bagchi:2009my, Russian}. In the introduction, we had remarked that in four dimensions, if the BMS group is not extended to include all conformal transformations, then it consists of the semi-direct product of the global conformal group in two dimensions and the supertranslations. The situation is similar in three dimensions. We can look at the ``restricted'' BMS algebra, with only the global part of the conformal transformations included. This contains $L_{\{0, \pm 1 \}}$ together with all the $M_n$'s. So, yet another curious observation of the BMS/GCA correspondence is that the features of the ``restricted'' BMS group in 3 dimensions is encoded in the symmetries of the Euler equations in $1+1$ dimensions.

\subsection{Central Charges}

Now, let us look at the realizations of central charges on both sides of the correspondence. 
In the gravity side, a Brown-Henneaux like analysis enables one to write down a central charge 
\cite{Barnich:2006av}
\be{central}
\mathcal{K}_{\xi,\xi^\prime} = \frac{1}{8\pi G}\int_0^{2\pi} d\theta
 \left[ \p_\theta Y^\theta (\p_\theta\p_\theta T^\prime + T^\prime) -
\p_\theta Y^{\prime\theta} (\p_\theta\p_\theta T + T)
 \right]  .
\ee
In terms of the generators $\mathcal{Q}_{P_n} = \cP_n,
\mathcal{Q}_{J_n} = \cJ_n$, we get the centrally extended algebra
\ben{bms_charge}
[\cJ_m,\cJ_n ] = (m-n) \cJ_{m+n},\qquad
[\cP_m,\cP_n ] = 0,\crcr
[\cJ_m,\cP_n] = (m-n) \cP_{m+n}+\frac{1}{4G} m(m^2-1)\delta_{n+m,0}.
\een
The central charge cannot be absorbed into a redefinition of the generators. Only the
commutators of generators involving either $\cJ_0,\cJ_1, \cJ_{-1}$ or
$\cP_0,\cP_1,\cP_{-1}$ corresponding to the exact Killing vectors of
the Poincar\'e algebra $\mathfrak{iso(2,1)}$ are free of central
extensions.

The discussion of the GCA was at the classical level of vector fields. At the quantum level the two copies of the Virasoro get respective central extensions
\ben{relalg}
[\L_m, \L_n] &=& (m-n) \L_{m+n} + {c \over 12} m(m^2-1) \delta_{m+n,0}\,, 
\nonumber \\[1mm]
[\bL_m, \bL_n] &=& (m-n) \bL_{m+n} 
+ {\bar c \over 12} m(m^2-1)\delta_{m+n,0}\,.
\een 
Considering the linear combinations \eq{GCArepn} which give rise to the GCA generators as in \eq{Vir2GCA}, we find 
\ben{gcawc}
[L_{m}, L_{n}] &=& (m-n) L_{m+n} + C_1 m(m^2-1) \delta_{m+n,0}\,, \crcr 
[L_{m}, M_{n}] &=& (m-n) M_{m+n} + C_2 m(m^2-1) \delta_{m+n,0}\,, \crcr 
[M_{m}, M_{n}] &=& 0\,.
\een
This is the centrally extended GCA in 2d.
Note that the relation between central charges is 
\be{centch}
C_1 = {{c+\bar c} \over 12}\,, \qquad {C_2 \over \e} = {{\bar c-c} \over 12}\,.
\ee
Thus, for a non-zero $C_2$ in the limit $\e\rightarrow 0$ we see that we need $\bar c-c \propto
\O({1\over \e})$. At the same time requiring $C_1$ to be finite we find
that  $c+\bar c$ should be $\O(1)$. 

So, matching the central charges on both sides we get 
\be{gr-cc}
C_1=0, \quad C_2 = {1 \over 4G}
\ee 

\subsection{BMS from AdS: Analogies and differences}

We would like to compare and contrast our approach in \cite{Bagchi:2009} in obtaining the 2d GCA from the relativistic Virasoro with the approach of \cite{Barnich:2006av} in deriving the $\mathfrak{bms_3}$ from Anti-de Sitter space. 
Let us remind ourselves of the construction of \cite{Barnich:2006av}. 
The algebra~\eqref{bms_charge} has many features in common with the
anti-de Sitter case: it has the same number of generators, and a
Virasoro type central charge. If one introduces the negative cosmological constant $\Lambda =
-\frac{1}{R^2}$ and considers
\be{alg_bms2}
[ {J_m},{J_n}] = (m-n) J_{m+n},\quad [ {P_m},{P_n}] = \frac{1}{R^2} (m-n)J_{m+n},\quad [{J_m},{P_n}
] = (m-n) P_{m+n},
\ee 
the $\mathfrak{bms}_3$ algebra~\eqref{alg_bms} corresponds to the case
$R\rightarrow \infty$. For finite $R$, the charges
$\L_m^\pm$ corresponding to the generators $L_m^{\pm} = {1\over2} (\, R
P_{\pm m} \pm J_{\pm m})$ form the standard two copies of the
Virasoro algebra,
\begin{eqnarray}
 [\L^\pm_m,\L^\pm_n \} &=&
(m-n) \L^\pm_{m+n}
+\frac{c}{12}m(m^2-1)\delta_{n+m,0} ,\qquad
\{ \L^\pm_m,\L^\mp_n \} =\, 0,
\end{eqnarray}
where $c= \frac{3R}{2G}$ is the central charge for the anti-de Sitter case.

This construction seems almost identical to the one described for the GCA previously. But there are some subtle differences. One notices that the linear combinations in the two cases are not the same, with 
\be{Vir2BMS}
P_m = {1\over R} (\L_m + \bL_{-m}), \quad J_m = (\L_m - \bL_{-m})
\ee  
as opposed to \refb{Vir2GCA}. This also sheds light on the reason why there is a central term in the $[P,J]$ commutator as opposed to the $[J,J]$. If the central terms were different for the left and right movers ($c^{\pm}$), the $[P,J]$ would have a $c^+ + c^-$ term while $[J,J]$ would have a $c^+ - c^-$ term. Here, in the case discussed above $c^+ = c^-$ and hence we get just the one central term. This corresponds to the $c = -\bar c$ in the GCA case. Notice that like $C_2$ in the GCA case, the central term needs to be very large to avoid the contraction and hence $c= \frac{3R}{2G}$.

\section{A BMN route to BMS}

In \cite{Bagchi:2009my}, we proposed a gravity dual of the GCA by taking a parametric limit of the bulk $AdS_{d+2}$ geometry. 
Consider the metric of $AdS_{d+2}$ in Poincare coordinates
\be{poinmet} 
ds^2 = {1\over z^{\prime 2}}(\eta_{\mu\nu}dx^{\mu}dx^{\nu}-dz^{\prime 2})={1\over z^{\prime 2}} (dt^{\prime 2} - dz^{\prime 2} -d x_i^2)
\ee 
The nonrelativistic scaling limit that was considered was 
\be{adsnrelscal}
t^{\prime} ,z^{\prime} \rightarrow t^{\prime} , z^{\prime} \qquad   x_i \rightarrow \e x_i.
\ee
The scaling of $t$ and $x_i$ were motivated by the boundary scaling.  Since the radial direction of the $AdS_{d+2}$ is an additional  dimension, we need to fix its scaling. The radial direction is a measure of the energy scales in the boundary theory via the usual holographic correspondence. We therefore expect it to also scale like time i.e. as  $\epsilon^0$. This means that in the bulk the time and radial directions of the metric {\it both} survive when performing the scaling. Together these constitute an $AdS_2$ sitting inside the original $AdS_{d+2}$. 
The bulk dual was proposed to be a Newton-Cartan like $AdS_2 \times R^d$ where the metric was degenerate and the dynamical quantities were the non-metric connections. The GCA in the bulk was shown to emerge by taking this limit on the Killing vectors of $AdS_{d+2}$. In what follows, we suggest a different bulk realization of the GCA, one relevant to the asymptotically flat space realization we have discussed earlier. We will show that a scaling exactly similar to \refb{adsnrelscal} gives rise to flat space metric from the AdS metric. 

Let us concentrate on $AdS_3$. We will now re-introduce factors of the AdS radius $R$. We would take a Penrose limit of the AdS metric in the co-ordinates stated above\footnote{This was obtained in collaboration with Rajesh Gopakumar.}. The Poincare patch has a horizon at $z^{\prime}=\infty$ and to extend the coordinates beyond this we will choose to follow an infalling null geodesic, in an analogue of the Eddington-Finkelstein coordinates. Therefore define $z=z^{\prime}$ and $t=t^{\prime}+z^{\prime}$. 
In these coordinates
\be{adsefink}
ds^2 = {1\over z^{2}} (-2dt dz+dt^{2})={dt\over z^{2}}(dt-2dz).
\ee 
Let's consider the bulk metric with the radius included. 
\be{blkmet}
ds^2={R^2\over z^{2}}(-dt(2dz-dt)-dx^{2}) 
\ee
This will give a non-degenerate metric only if we have the scaling 
\be{pnrs}
x={\mu \over R} x, \ \ \ \ t,z \sim \O(1), \ \ \ t-2z ={\mu^2\over R^2}v
\ee
with $R\rightarrow \infty$ and keeping $\mu, v, x_i, t$ finite. 
The resulting metric is
\be{}
ds^2={4\mu^2\over t^{2}}(dt dv-dx^{2})
\ee
where we have kept the leading order terms as $R\rightarrow \infty$. (Notice $z^2 = t^2 + \O(R^{-2})$ and hence the replacement.)
This is like a BMN limit \cite{BMN} where we are zooming into the vicinity of the null radial geodesic. Note that
$t-2z=t'-z'$ in terms of the original Poincare coordinates that we started out with. 

However, this metric is actually flat. This can be seen by writing $x=t\rho$. This gives the metric
\be{flat}
{ds^2\over 4\mu^2}={dz\over t^2}(dv-\rho^2dt-2\rho td\rho)-d\rho^2 
=-d({1\over t})d(v-\rho^2t)-d\rho^2.
\ee
which we see is clearly a flat metric on $R^{2,1}$ when we define $\tilde{u}={1\over t}$ and
$\tilde{v}=v-\rho^2 t$.
We have kept only the leading terms in the above computation. 

The above computation shows that by taking a non-relativistic limit \refb{pnrs} which is almost exactly like \refb{adsnrelscal}, with an additional condition on $t-z$, we can recover a flat space. It was shown in \cite{Bagchi:2009my}, that the Killing vectors of AdS in the limit \refb{adsnrelscal} give rise to an infinite algebra in the bulk which precisely reduces to the GCA on the boundary and satisfy the same commutation relations in the bulk. Given the connection between $\mathfrak{bms_3}$ and $\mathfrak{gca_2}$, it is satisfying that one has been able to recover a flat space metric using the same limit. 

In terms of the $AdS_3/CFT_2$ correspondence, the above mentioned $BMS_3/GCA_2$ is thus a limit where on the gravity side one takes the radius of AdS to infinity while on the field theory side, one takes the speed of light to infinity. So, this seems to indicate an equivalence between the radius of AdS and the speed of light in the CFT.

\section{The $\mathfrak{bms_4 / gca^{s=1}_{3}}$ correspondence}

\subsection{The BMS group in 4 dimensions}

The structure of the BMS group in four dimensions as before is dictated by the structure of the spacetime at null infinity, which is now $S^2 \times R$. As in the case of the three dimensional BMS group, if one does not want to restrict to 
globally well defined transformations on the two-sphere, we get two copies of the Witt algebra. The general solution to the conformal Killing equations is
$Y^\zeta=Y(\zeta)$, $Y^{\bar\zeta}=\bar Y(\bar\zeta)$, with $Y$ and
$\bar Y$ independent functions of their arguments. The standard basis
vectors are chosen as
\be{eq:55}
l_n=-\zeta^{n+1}\frac{\p}{\p\zeta},\quad \bar l_n=-\bar
\zeta^{n+1}\frac{\p}{\p\bar \zeta},\quad n\in \mathbb Z
\ee
At the same time, let us choose to expand the generators of the
supertranslations in terms of
\be{eq:15}
  T_{m,n}=P^{-1}\zeta^m\bar\zeta^n, 
\quad m,n\in\mathbb Z. \quad (P(\zeta,\bar\zeta)={1\over 2}(1+\zeta\bar\zeta))
\ee 
In terms of the basis vector $l_l\equiv (l_l,0)$ and
$T_{mn}=(0,T_{mn})$, the commutation relations for the complexified
$\mathfrak{bms}_4$ algebra read
\ben{bms4}
[l_m,l_n]=(m-n)l_{m+n},\quad [\bar l_m,\bar l_n]=(m-n)\bar l_{m+n},\quad [l_m,\bar l_n]=0, \crcr 
[l_l,T_{m,n}]=(\frac{l+1}{2}-m)T_{m+l,n},
\quad [\bar l_l,T_{m,n}]= (\frac{l+1}{2}-n)T_{m,n+l}. 
\een
Two copies of the Witt algebra indicate that we would need to look beyond usual GCAs in any dimensions as the field theory realizations of this symmetry.

\subsection{Semi-Galilean Conformal Algebras}

In order to find a field theoretic description of the above symmetry, in this section we study non-relativistic limit of relativistic
conformal algebra in $d+1$ dimensions by making use of a general contraction \cite{Alishahiha:2009np}. 
\be{scale} 
t\rightarrow t,\quad y_\alpha\rightarrow
y_\alpha,\quad x_i\rightarrow \epsilon x_i, 
\ee 
where $\alpha=1,\cdots, s$ and $i=s+1,\cdots d$. The contraction is
defined by the above scaling in the limit of $\epsilon\rightarrow
0$. 

We start from a CFT in $d+1$ dimensions. The GCA was a specific example of the semi-GCA with $s=0$. We label semi-GCAs by  $\mathfrak{gca^{s}_{n}}$. (In our notation, GCA is $\mathfrak{gca^{s=0}}$. Let us consider the case of $s=1$. As in the $s=0$ case, there is a finite algebra which is obtained by contraction and then this can be given an infinite dimensional lift. It is useful to define new coordinates $u=t+y,\;v=t-y$. The infinite generators are \cite{Alishahiha:2009np}
\ben{Infboun}
L_n=u^{n+1}\partial_u+\frac{n+1}{2}u^nx_i\partial_i,\quad
\bar{L}_n=v^{n+1}\partial_v+\frac{n+1}{2}v^nx_i\partial_i, \\
M_{i\;rs}=-u^rv^s\partial_i. \quad
J_{ij\;nm}=-u^nv^m(x_i\partial_j-x_j\partial_i), 
\een 
one finds an infinite dimensional algebra as follows 
\ben{}
&&[L_n,L_m]=(n-m)L_{n+m},\qquad [\bar{L}_n,\bar{L}_m]=(n-m)\bar{L}_{n+m},\cr &&\cr
&&[M_{i\;nm},M_{j\;n'm'}]=0,\qquad[L_n,\bar{L}_m]=0,
\cr &&\cr
&&[L_n,M_{i\;ml}]=\left(\frac{n+1}{2}-m\right)M_{i\;(n+m)l},\;
\quad [\bar{L}_n,M_{i\;ml}]=\left(\frac{n+1}{2}-l\right)M_{i\;m
(n+l)} \cr &&\cr
&&[L_n,J_{ij\;ml}]=-mJ_{ij\;(n+m)l},\qquad
[\bar{L}_n,J_{ij\;ml}]=-lJ_{ij\;m (n+l)}\cr &&\cr
&&[M_{l\;nm},J_{ij\;n'm'}]=\left(\delta_{jl}M_{i\;(n+n')(m+m')}-\delta_{il}M_{j\;(n+n')(m+m')}\right).
\een
$J_{ij\;nm}$'s generate an $so(d-1)$ affine algebra.

We are interested in the case where the dimension of the field theory is three and the co-ordinates are $t,y,x$, viz. one of each kind mentioned above. Then the additional vector indices drop off, as do the generators of rotation. Then it is straight forward to observe that this restricted algebra which we call $\mathfrak{gca^{s=1}_{3}}$ is isomorphic to \refb{bms4}.
So we see that in this case we have a correspondence between asymptotically flat four dimensional space described by the $\mathfrak{bms_4}$ algebra and a conformal field theory in three dimensions where one of the spatial dimensions is non-relativistic.

\section{Remarks on a general correspondence}

The BMS algebra is infinite dimensional in 3 and 4 dimensions as the group of isometries of the circle and the sphere are enhanced to the full conformal algebra. Even in the case of the GCA and its cousin, called the semi-GCA with $s=1$ we have seen similar enhancements. For $s=2$ and beyond, we don't get this enhancement. A way of understanding this fact is to remember the gravity description in terms of the Newton-Cartan structure, as defined in \cite{Bagchi:2009my}. The GCA had a bulk piece with an $AdS_2$ and the s=1 GCA had a similar $AdS_3$ counterpart. The higher sGCA would have $AdS_{2+s}$ factors in the Newton-Cartan structure, but these would not have enhanced Virasoro like symmetries \cite{Alishahiha:2009np}. This suggests a general $BMS_{3+s}/sGCA_{2+s}$ correspondence. In words, we should keep one non-relativistic direction in the conformal field theory that is dual to the one higher dimensional BMS group. For example, when we are looking at the BMS algebra in five dimensions, we should look at a four-dimensional $s=2$ GCA. 

Let us comment briefly on this above correspondence in the context of realizing it as a limit of an $AdS/CFT$ correspondence, like we saw in the three dimensional case. One should easily be able to obtain the flat space limit of the respective $AdS$ space by looking at higher dimensional analogues of the radially infalling null-ray and scaling the co-ordinates in a manner described in Sec.~4. The point to note is that now all differences between relativistic co-ordinates would scale as $R^{-2}$. The speed of light in the non-relativistic direction of the field theory would be the dual to the AdS radius in the bulk theory. There is however one point that needs more understanding. In the three-dimensional case ($\mathfrak{bms_3}$), the fact that the algebra is infinite came from the existence of the two copies of the parent Virasoro algebra of $AdS_3$. For $\mathfrak{bms}_n$ where $n > 4$, there is no enhancement of the algebra and this is in keeping with the parent $AdS_n$ theory. But the case of $n=4$ is curious from this point of view as there is no infinite algebra of asymptotic isometries of $AdS_4$ but we still get an infinite algebra in the limit\footnote{This is somewhat analogous to the appearance of the infinite $\mathfrak{gca}_d^{s=0}$ from the relativistic conformal algebra in $d>2$.}. We would like to understand this better.

\section{Conclusions}
In this note, we have looked at two seemingly unrelated pictures, that of asymptotically flat spaces and non-relativistic conformal systems in one lower dimension and shown that they are equivalent at the level of the symmetry algebras. We have pointed out explicitly the cases of three and four dimensions and made some general remarks about a correspondence in all dimension of space-time. 

We have managed to understand the case of three dimensional gravity better by looking at the theory in Anti de Sitter space. A BMN like non-relativistic limit on radial infalling null ray gave us a flat metric and a similar non-relativistic limit on the Killing vectors in \cite{Bagchi:2009my} had revealed the bulk GCA, which is the BMS in three dimensions. On the boundary, we had taken a similar limit to obtain the GCA in two dimensions. This seems to indicate that the asymptotically flat space limit of $AdS_3$ is equivalent to looking at the non-relativistic limit of the conformal algebra in two dimensions and a duality between the radius of AdS and the speed of light in the boundary conformal theory. We should stress that as of now, we do not fully understand how to reconcile the seemingly different space-time actions of the BMS and the GCA. This is a point that we are looking to address in the near future.
 
We have looked at the structure of the non-relativistic field theory in some detail in \cite{Bagchi:2009}. It was found there that most of the answers in the GCA can be obtained in a spirit very similar to the techniques of 2-d conformal symmetry. Even though the mixing of the holomorphic and anti-holomorphic components make life somewhat difficult, the calculations are not intractable. Given this correspondence between flat space and the GCA, it is tempting to ponder on the consequences of our analysis in \cite{Bagchi:2009} and hope to make statements about answers on the gravity side. For this, a first step would be to identify the parameters on both sides. We had labelled the GCA with the eigenvalues of boosts and dilatations in \cite{Bagchi:2009ca, Bagchi:2009}. It would be useful to understand the relations of these to physical quantities in the asymptotically flat spaces. The correlation functions of the GCA were also computed in \cite{Bagchi:2009ca, Bagchi:2009}. The correlation functions in the field theory would map to onshell amplitudes in gravity. It is plausible that one would be able to make statements about the S-matrix in asymptotically flat spacetimes by using the techniques of the GCA. We are unaware of any such analysis using the BMS algebra and this is an avenue definitely worth exploring. 

The four dimensional case, i.e. $BMS_4/sGCA_3$ is a case we have talked less about in this note. This is however the more interesting map for physical systems because of the obvious reason that we are talking about asymptotically flat four dimensional space. On the field theory side, this is a case which has been far less studied. One must look at the representations and in a spirit similar to \cite{Bagchi:2009ca}, one should be able to construct the Hilbert space and find the two and three point functions by looking at just the global part of the algebra. One crucial point of difference that can be observed immediately is that the boosts would not commute with the $L_0$ and $\bar{L}_0$. So we would have significant differences with $s=0$ GCA. 

This note is a first step in the direction of using the non-relativistic conformal techniques to study the holography of flat space, which we hope would be a worthwhile exercise.

\subsection*{Acknowledgements}
It is a pleasure to acknowledge discussions with L. Alday, S. Rey, Y. Tachikawa, A. Zhiboedov and thank S. Banerjee, R. Gopakumar, D. Ghoshal and A. Sen for valuable comments on the manuscript. The author would especially like to thank Juan Maldacena for many illuminating conversations. The author would also like to acknowledge the warm hospitality of the Institute of Advanced Studies, Princeton where the research was carried out.


\end{document}